\documentclass[11pt,twoside]{article}
\usepackage{./asp2014}

\usepackage{graphicx}

\aspSuppressVolSlug
\resetcounters

\begin{document}

\title{The IRON Project: Photoionization of Fe ions}
\author{Sultana N Nahar}
\affil{The Ohio State University, Columbus, OH 43210, USA; 
\email{nahar.1@osu.edu}}

\paperauthor{Sample~Author1}{Author1Email@email.edu}{ORCID_Or_Blank}{Author1 Institution}{Author1 Department}{City}{State/Province}{Postal Code}{Country}
\paperauthor{Sample~Author2}{Author2Email@email.edu}{ORCID_Or_Blank}{Author2 Institution}{Author2 Department}{City}{State/Province}{Postal Code}{Country}
\paperauthor{Sample~Author3}{Author3Email@email.edu}{ORCID_Or_Blank}{Author3 Institution}{Author3 Department}{City}{State/Province}{Postal Code}{Country}

\begin{abstract}
The IRON Project, initiated in 1991, aims at two main objectives, i) study
the characteristics of and calculate large-scale high accuracy data for
atomic radiative and collisional processes, and ii) application in solving
astrophysical problems. It focuses on the complex iron and iron-peak
elements commonly observed in the spectra of astrophysical plasmas. The
present report will illustrate the characteristics of the dominant atomic
process of photoionization that have been established under the project and the
preceding the Opacity Project and their importance in applications.
\end{abstract}

\section{Introduction: Photoionization and the Opacity and Iron Projects}

Photoionization, photo-excitation, electron ion recombination, electron impact 
excitation and ionization are the most dominant atomic processes in 
astrophysical plasmas. Modeling of astrophysical spectra for abundances, 
ionization fractions, diagnostics for the physical and chemical conditions, 
plasma opacity require the cross sections for photoionization when the
plasma is around or near a radiative source, such as, a star. 

Photoionization process of an ion of charge $+z$ can be direct:
\begin{equation}
X^{+z} + h\nu \rightleftharpoons X^{+z+1} + \epsilon
\end{equation}
This gives smooth background cross section. The process can be in two steps 
when an intermediate doubly excited state is formed:
\begin{equation}
{ X^{+z} + h\nu \rightleftharpoons (X^{+z})^{**}
 \rightleftharpoons X^{+z+1} + \epsilon}
\end{equation}
In collisional description, an electron colliding with the ion can form 
an intermediate state where two electrons are excited. It is called
an autoionizing state if it lies above the ionization threshold and breaks 
down either to autoionization (AI) when the electron goes free in a 
radiation-less process as the inner electron drops to the ground state or 
to dielectronic 
recombination (DR) when electron is captured by emission of a photon,
\begin{equation}
 e  + X^{+z} \leftrightarrow (X^{+z-1})^{**} 
 \leftrightarrow \left\{ 
 \begin{array}{ll}
  e + X^{+z} & \mbox{AI} \\ X^{+z-1} + h\nu & \mbox{DR}\end{array}
 \right\}
\end{equation}
DR is the inverse of photoionization. 
The autoionizing state is a quantum state of energy 
\begin{equation}
{E_xnl = E(X^{+z}\nu l)^{**} = E_x-\frac{z^2}{\nu^2} =  E_x-\frac{z^2}
{(n-\mu)^2} }
\end{equation}
where $Ex$ is an excited state of the residual or core ion relative to its
ground state, $nl$ is the quantum level of the outer electron. $n$ is 
basically the effective quantum number $\nu = n-\mu$ where $\mu$ is the
quantum defect describing the departure from a hydrogenic state. 
An autoionizing state introduces a resonance in photoionization cross 
section ($\sigma_{PI}$). Levels $E_xnl$ with different $nl$ below an excited 
threshold $Ex$ form the Rydberg series. 
So to each excited state of the residual ion, there can be a series of 
Rydberg resonances which show similar structure. A resonance is different 
from the typical rise in the background cross section at an excited threshold 
energy of an inner shell of the core ion.
Identification of resonances can be complex for overlapping Rydberg states
belonging to different $E_x$ lying close together where the interference 
can modify the structures. Also depending on the values of the 
excitation rate coefficients and interference the strength of the resonances 
may reduce. In addition Seaton resonances (Yan and Seaton 1987), different 
from a Rydberg resonance can form (discussed later).

The autoionization rate ($A_a$)is high, about 10$^{15-16}s^{-1}$, and remains 
the same for all effective quantum number of the Rydberg series of resonances. 
The radiative decay rate ($A_r$), about 10$^{8-12}s^{-1}$ for low lying 
levels, but increases as $\nu^3$ and becomes comparable to $A_a$ near and 
below the excited threshold of the residual ion. Both rates can be 
comparable at lower $\nu$ for highly charged He- and Li-like ions. 

The Opacity Project (OP) (Seaton 1987, Opacity Project 1995,1996) carried 
out the first systematic study of photoionization of the ground and many 
excited states of astrophysically important atoms and ions from H to Fe.
The atomic and opacity data computed are available in databases TOPbase 
(TOPbase) and OPServer (OPserver) respectively.
OP solved many longstanding astrophysical problems. However, recent advances 
in technology for experimental measurements and astrophysical
observations require data of higher accuracy. Follow-up of the OP
the IRON Project (IP) (Hummer et al 1993) includes relativistic fine structure
effect to achieve it in the study of the collisional and 
radiative processes of the much needed elements, and for the iron and 
iron peak elements. A significant amount of calculated data obtained under 
the IP are available online at TIPbase (TIPbase) and at atomic data page 
NORAD-Atomic-Data (NORAD) at the Ohio State University.

This report presents a review of the features in photoionization established
under the OP and the IP using iron ions as examples. These ions are being 
studied particularly to obtain accurate opacity in the sun. Opacity depends 
mainly on photoionization adn photoexcitation and gives a measure of 
radiation absorption in the plasma. It is related to the elemental 
abundances which has been a longstanding problem for the sun 
(e.g.  Asplund et al 2004, 2009, Bailey et al 2015, Nahar and Pradhan 2016).

\section{Theory}

The throretical method used for the study of photoionization is the R-matirx
method with close coupling (CC) approximation for the wavefunction expansion.
(e.g. Berrington et al 1987, {\it Atomic Astrophysics and Spectroscopy} (AAS)). 
CC approximation is the most precise way to generate the resonances naturally. 
A complete treatment of photoionization can be found in the textbook AAS.
A brief outline of the theory is given below for the general guidance of 
the readers. Under the OP and the IP the R-matrix method (e.g.  Burke and 
Robb 1975, AAS) was extended extensively for calculating excited states, 
oscillator strengths, photoionization cross sections $\sigma_{PI}$ (Seaton 
1987, Hummber et al 1993, Berrington et al 1987, 1995). The R-matrix package 
of codes under the IP include the relativistic fine structure effects in 
Breit-Pauli approximation giving the name of the method as Beit-Pauli 
R-matrix or BPRM method.

In the CC approximation the atomic system is represented by the 'target' 
or the 'core' ion with (N+1)$^{th}$ electron as the interacting electron. 
The (N+1)$^{th}$ electron may be bound (E$<$ 0) or in the continuum (E$\geq$0).
The total wave function, $\Psi_E$, in a $J\pi$ symmetry is expressed as
\begin{equation}
\Psi_E(\mbox{e+ion}) = A \sum_{i} \chi_{i}(\mbox{ion})\theta_{i} + \sum_{j} c_{j} \Phi_{j},
\end{equation}
where $\chi_{i}$ is wavefunction of the core ion at specific level
$S_iL_i(J_i)\pi_i$ coupled with the (N+1)$^{th}$ electron function
$\theta_{i}$. The first sum is over the ground and excited states of the
the core ion. $A$ is the anti-symmetrization operator. The (N+1)$^{th}$
electron with kinetic energy $k_{i}^{2}$ corresponds to a channel labeled
$S_iL_i(J_i)\pi_ik_{i}^{2}\ell_i(SL(J)\pi)$. The $\Phi_j$s are bound 
channel functions of the (N+1)-electrons system that account for short 
range correlation not considered in the first term and the orthogonality 
between the continuum and the bound electron orbitals of the core ion. It 
is assumed that the core ion orbitals remain the same before
and after ionization that is the system is described by a wavefunction
in spherical coordinates centered on the heavy point like and spinless
nucleus with electric charge number +$z$. The wavefunctions of the core ion
can be obtained from configuration interaction atomic structure calculations,
such as using the program SUPERSTRUCTURE (Eissner et al 1974, Nahar et al.
2003).

The relativistic Hamiltonian in the BPRM method is given by (e.g. AAS)
 \begin{equation}
 H_{N+1}^{\rm BP} = \sum_{i=1}\sp{N+1}\left\{-\nabla_i\sp 2 - \frac{2Z}{r_i}
 + \sum_{j>i}\sp{N+1} \frac{2}{r_{ij}}\right\}+
 H_{N+1}^{\rm mass} + H_{N+1}^{\rm Dar} + H_{N+1}^{\rm so}.
 \end{equation}
where the relativistic mass correction, Darwin and spin-orbit interaction
terms are
\begin{equation} 
H_{N+1}^{\rm mass} = -{\alpha^2\over 4}\sum_i{p_i^4},
~H_{N+1}^{\rm Dar} = {Z\alpha^2 \over 4}\sum_i{\nabla^2({1
\over r_i})}, 
~H_{N+1}^{\rm so}= Z\alpha^2 
\sum_i{{\bf l_i.s_i}\over r_i^3}.
\end{equation}
respectively. 
The Hamiltonian has additional two-body interactions terms (e.g. Nahar et al 
2003). The BPRM method includes part of their contributions and neglect 
of the magnetic interaction (i.e. mutual spin-orbit, spin-other-orbit and 
spin-spin) of the colliding/outer electron with the valence electrons of the 
core ion. 

Substitution of $\Psi_E(e+ion)$ in the Schrodinger equation
$H_{N+1}\mit\Psi_E = E\mit\Psi_E$
introduces a set of coupled equations. They are solved using the R-matrix
method where the space in divided into two regions, the inner region of a 
sphere of radius $r_a$ with the ion at the center and the outer 
region. The wavefunction inside is represented by an expansion. known as
the R-matrix basis. $r_a$, is chosen large enough for the electron-electron 
interaction potential is zero outside it.  Beyond $r_a$ the wavefunction 
is treated as Coulombic by the perturbation from the long-range electric 
multipole potentials arising from the target orbitals and elaborate angular 
algebra.

The transition matrix element for the dipole transition is given by
$<\Psi_B || {\bf D}|| \Psi_{F}>$,
where ${\bf D} = \sum_i{r_i}$ is the dipole operator and the sum is over the
number of electrons; $\Psi_B$ and $\Psi_{F}$ are the initial bound and
final continuum wave functions. The transition matrix element can be reduced
to generalized line strength as (e.g. AAS)
\begin{equation}
{\bf S}= |<\Psi_j||{\bf D}||\Psi_i>|^2 =
 \left|\left\langle{\mit\psi}_f
 \vert\sum_{j=1}^{N+1} r_j\vert
 {\mit\psi}_i\right\rangle\right|^2 \label{eq:SLe},
\end{equation}
where $\mit\Psi_i$ and $\mit\Psi_f$ are the initial and final state wave
functions. The photoionization cross section ($\sigma_{PI}$) is proportional
to the generalized line strength as,
$\sigma_{PI} = {4\pi^2 \over 3c}{1\over g_i}\omega{\bf S},$
where $g_i$ is the statistical weight factor and $\omega$
is the incident photon energy. The resonances are introduced by the
interference in the transition matrix element of the bound wavefunction
which includes core ion excitations and continuum wavefunction of the outer
electron. For highly charged ions, radiation damping of 
autoionizing resonances can be treated 
using a scheme on fitting the dipole matrix of the autoionizing resonance 
(Sakimoto et al 1990, Pradhan and Zhang 1997, Zhang et al 1999)

\section{Results and Discussions}

Study of characteristic features of photoionization is important to learn
the process and for precise applications. The central field or distorted 
wave approximations can be used to obtain the approximate background 
cross sections $\sigma_{PI}$ (e.g. Reilman and Manson 1979, Gu 2008).
The resonances computed separately were added to the background cross section 
(e.g. Simon et al 2010). The rate for the bound-free transition obtained 
from atomic structure rate can be broadened with a profile function to 
form the resonant shape. These can not demonstrate accurate features and
can introduce large uncertainty. The following subsections illustrate 
features of $\sigma_{PI}$ obtained from R-matrix method. Numerical notation
is for an ion, such as Fe I for neutral and Fe II for Fe$^+$.

\subsection{Resonances in a two electrons system}

 \begin{figure}
 \centering
\includegraphics[width=3.75in,height=2.0in,angle=0.]{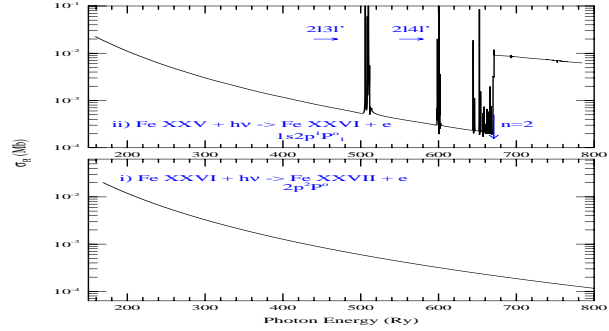}
\caption{ 
Photoionization cross sections $\sigma_{PI}$ illustrating \textbf{i)} smooth
decay of $2p^2P^o$ state of H-like Fe~XXVI and \textbf{ii)} 
resonant features in $1s2p^1P^o$ state of He-like Fe~XXV (Nahar et al 2001). 
The high peak resonances of type $2l3l'$, $2l4l'$ etc and enhancement in the 
background at the n=2 threshold at 670 Ry have been introduced by the dipole 
allowed transition 1s-2p in the core ion.}
\end{figure}
A hydrogenic ion has no core electron and hence can not form a doubly excited 
autoionizing state for a resonance. The $\sigma_{PI}$ of $2p(^2P^o)$ 
state of H-like iron ion, Fe~XXVI, in Figure 1 (i) shows smooth 
decay with photon energy (Nahar et al 2001). The $\sigma_{PI}$ for 
the $1s2p(^1P^o)$ state of the He-like iron ion, Fe~XXV (Figure 1 ii),
decays like a hydrogenic ion at lower energy (shown 
beyond the ionization threshold energy 159 Ry to elaborate the 
resonances), but exhibits strong high peak resonances of type $2l3l'$, $2l4l'$ 
etc belonging to n=2 states of the core ion (Nahar et al 2001). Considerable 
enhancement is seen at the excited n=2 threshold energy of about 670 Ry 
(pointed by an arrow) due to dipole allowed $1s-2p$ transition 
in the core ion. These features have been generated by the BPRM 
method using a 10-CC wavefunction expansion or 10 levels of the core ion 
which can be excited up to 4f level. These structures were not found in the 
earlier work using a 1-state calculations under the OP (Seaton in TOPbase).


\subsection{Resonances in ions with large number of electrons}

 \begin{figure}
 \includegraphics[width=2.50in,height=1.8in,angle=0.]{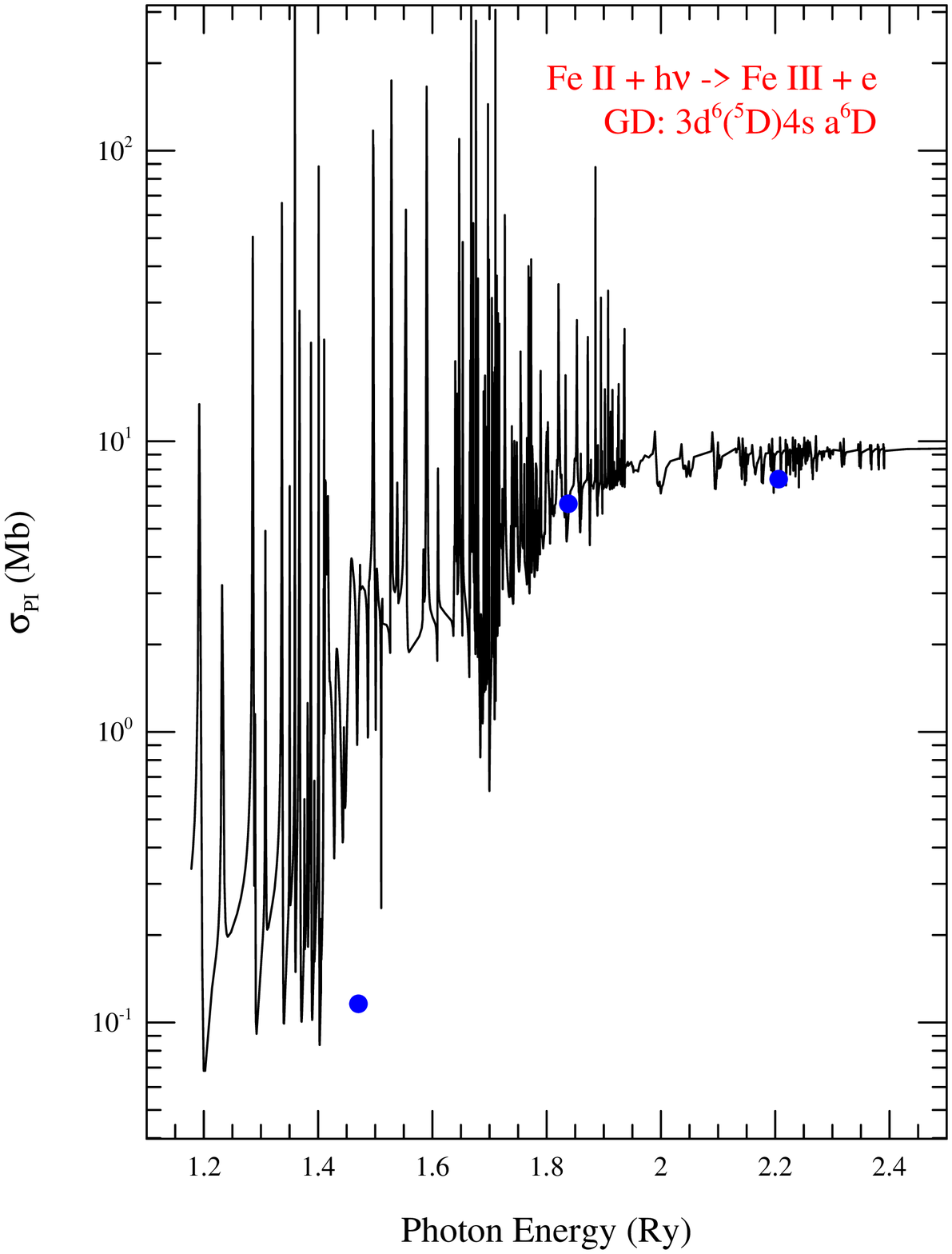}
 \includegraphics[width=2.75in,height=2.5in]{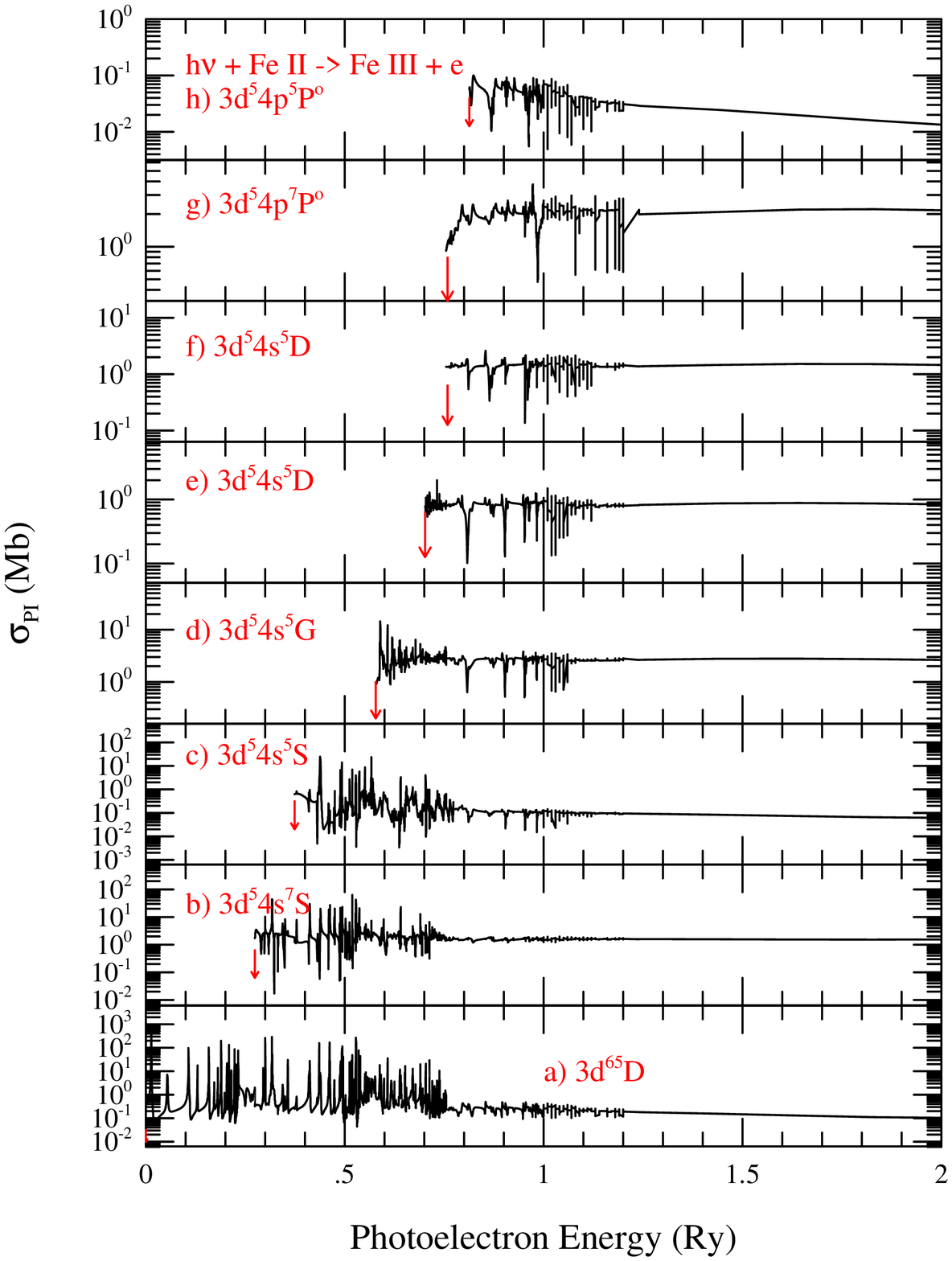}
\caption{Left: $\sigma_{PI}$ of the ground state, $3p^64s(^6D)$, of Fe~II
demonstrating extensive resonances in the low energy region due to strong 
electron-electron interaction among 25 electrons (Nahar and Pradhan 1994). 
The blue points are from central field approximation (Reilman and
Manson 1979). Right: Partial $\sigma_{PI}$ of the ground state
$3p^64s(^6D)$ for leaving the core ion in the ground (a) and 7 other excited 
states (b-h, thresholds pointed by arrows) as specified in the panels.
}
\end{figure}
Increasing number of electrons in the core ion usually increases number of 
excitations in the lower energy region and hence the number of Rydberg series 
of autoionizing resonances near the ionization threshold. These resonances 
are important for low to medium high temperature plasmas for integrated 
quantities containing an exponential factor such as recombination 
rates.
Seaton resonances (shown later) are usually more prominent in the high energy
and has more impact on high temperature plasmas.

Figure 2 (left panel) shows $\sigma_{PI}$ of the ground state, $3d^64s(^6D)$, 
of Fe~II obtained from the largest atomic 
calculations at that time (Nahar and Pradhan 1994). The wavefunction expansion 
included 83 states from n=2,3 complexes of the core ion Fe~III. 
The ion shows strong electron-electron correlation interaction effect by 
generating extensive resonances near the ionization threshold and over the 
energy range of about 1.3 Ry inclusive of the 83 states. The earlier 
$\sigma_{PI}$ obtained under the OP (Sawey and Berrington 1992) had 
incomplete correlation interaction and produced only the general background 
which is much lower than the present values.

The total $\sigma_{PI}$, such as that of Fe~II above, is obtained by summing 
up the partial $\sigma_{PI}$ corresponding to ionization leaving the residual 
ion at the ground and various excited states. With increase of photon energy
the ion is ionized with higher excitations of the core ion. Right panel of 
Figure 2  presents partial $\sigma_{PI}$ with resonant structures leaving 
the residual ion at the ground (a) and 7 (b-h) other lower excited states 
(Nahar and Pradhan 1994). The partial cross sections help in identification 
of the resonances.

\subsection{Photoionization of excited equivalent electron states}

The ground state photoionization typically has less features than those 
of excited states. (Fe~II in Figure 2 is an exception) and slow 
decay of the background cross section. The other states that show
slow decay of the background $\sigma_{PI}$ are the equivalent electron states. 
However, they have more resonant features in the lower energy region. These 
states are important contributor to processes relevant to it, such, 
electron-ion recombination. Figure 3 presenting $\sigma_{PI}$ of two equivalent 
electron states of Fe IV (Nahar 2005) obtained using a 16-CC wavefunction
expansion (Nahar and Pradhan 2005) shows the typical features of 
equivalent electron states. Narrow Rydberg resonances belong to various low 
lying excited states of the core ion form prominently at and near the 
ionization threshold and the background continues to remain high 
beyond the highest excitation threshold of the core ion included in the 
wavefunction expansion. Equivalent electron states do not show the wide 
Seaton resonances (illustrated belows)
 \begin{figure}
 \centering
 \includegraphics[width=4.0in,height=2.00in]{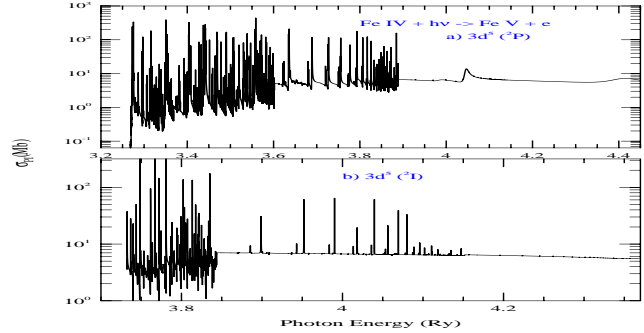}
\caption{
$\sigma_{PI}$ of two equivalent states $3d^5(^2P)$ and $3d^5(^2D)$ of Fe IV
(Nahar 2005) demonstrating continuing dominance of the background 
at higher energies.
}
\end{figure}

\subsection{Seaton Resonances due to PEC}

A Seaton resonance is different from a Rydberg resonance. It forms in 
$\sigma_{PI}$ of an excited state "i" with a single valence electron. The 
core ion at ground state absorbs the photon for photo-excitation-of-core (PEC) 
to a dipole allowed state of transition of energy $Ex$. The outer electron 
remains inactive temporarily or a "spectator" during the transition but leads 
to ionization as the excited core ion drops to the ground state. PEC interferes 
with the Rydberg resonances and manifests as a wider resonance with enhanced 
background, often by orders of magnitude, at an energy of $Ex-Ei$ from the 
ionization threshold energy $Ei$ of the state "i". 
Figure 4 illustrates characteristics of a Seaton resonance (pointed by arrow) 
in $\sigma_{PI}$ (Nahar 1996) - of the excited $3d^5(^6S)7p(^5P^o)$ state 
(left) and of a number of excited states of Fe~III (right). A Seaton resonance
is more distinct in higher excited states Figure 4 (right, b-f). Interference 
from electron-electron correlation of low lying excited states can reduce 
the prominence of the structure and may appear slightly shifted Figure 4 
(right, a,b). Seaton resonances cause the high energy behavior of $\sigma_{PI}$ 
non-hydrogenic for low to fairly high lying excited states contradicting 
their assumed hydrogenic $1/\nu^3$ behavior with energy. 
  \begin{figure}
  \centering
  \includegraphics[width=2.40in,height=1.50in]{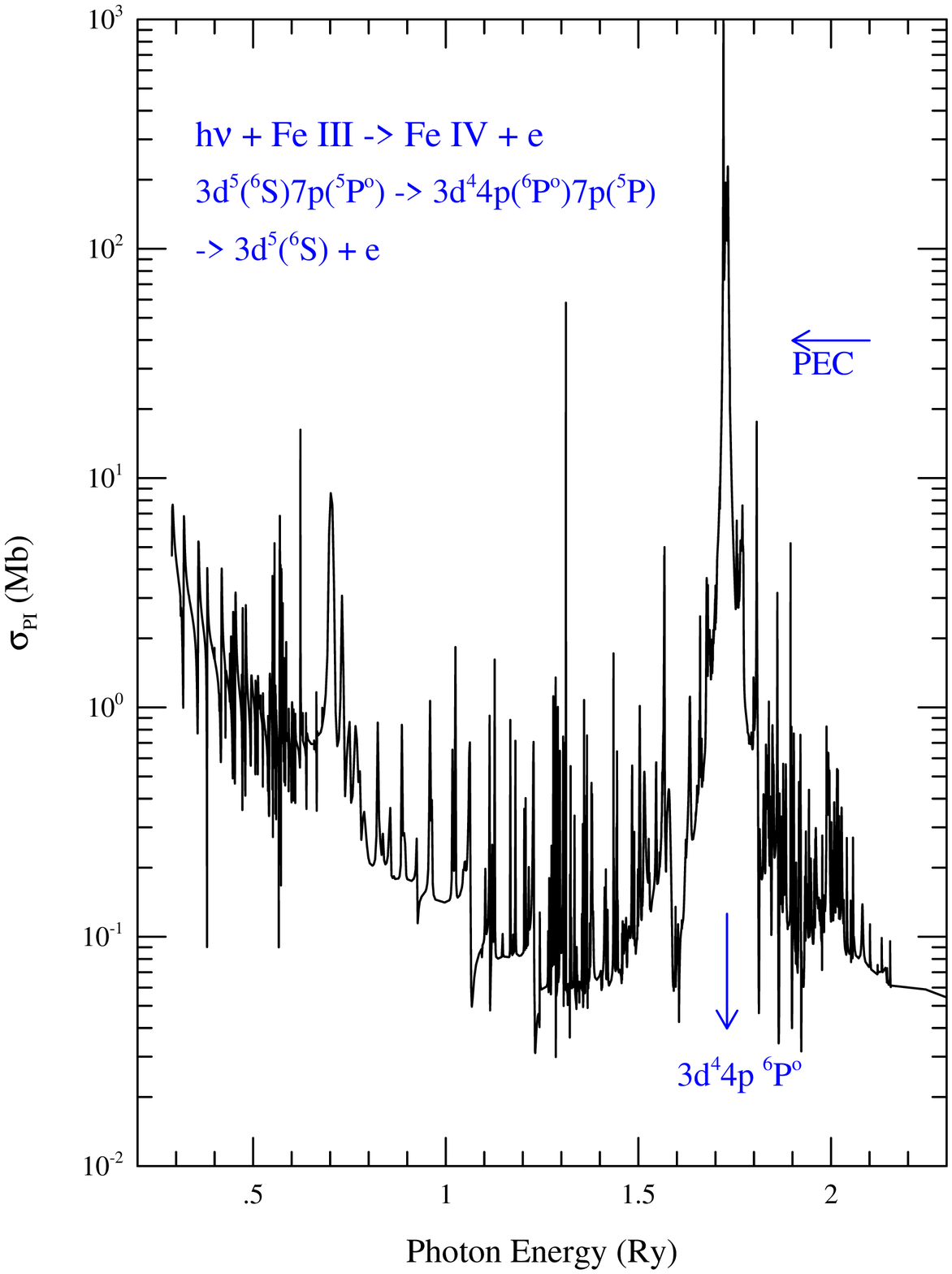}
  \includegraphics[width=2.55in,height=2.00in]{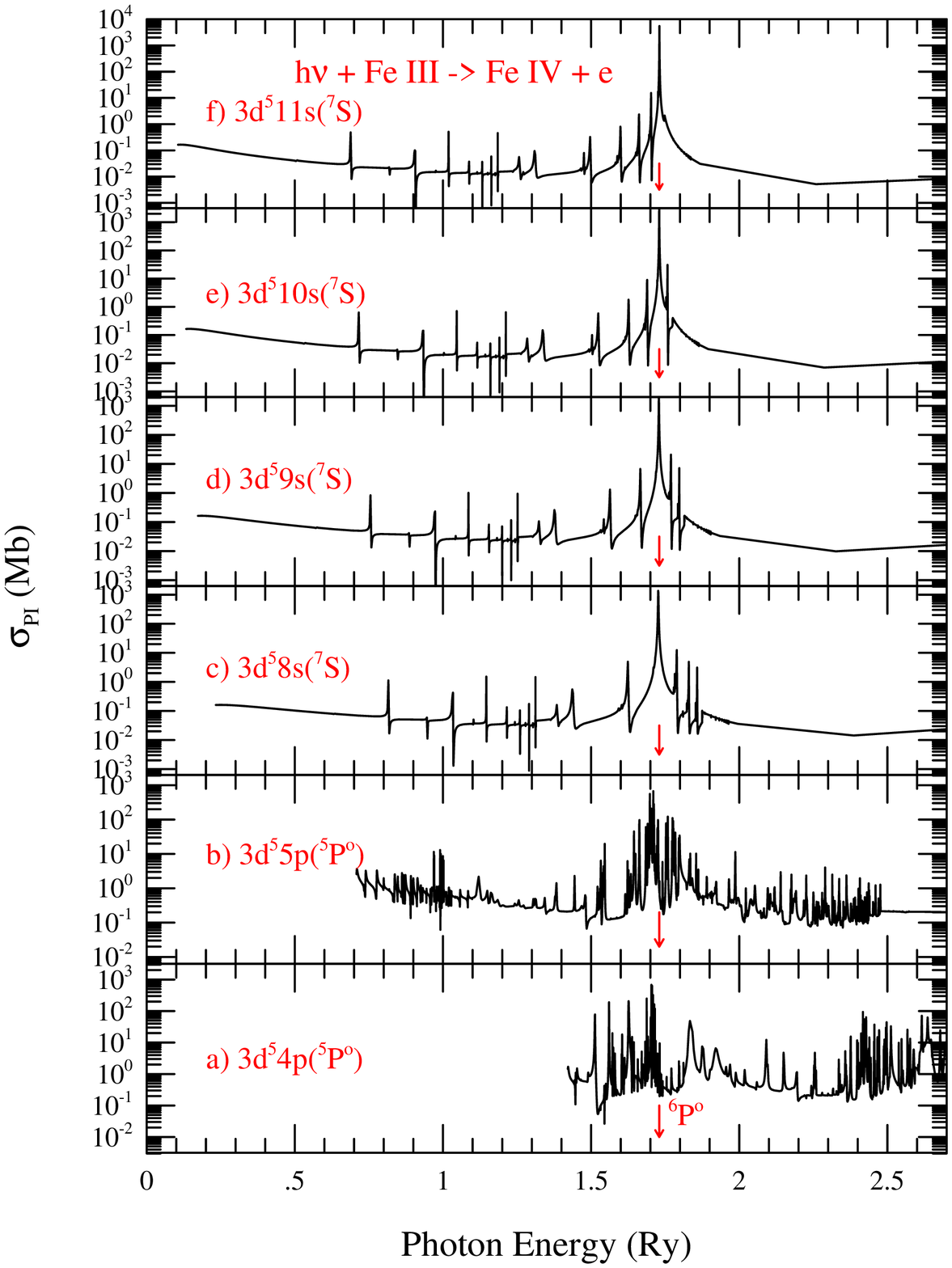}
 \caption{
 Left: $\sigma_{PI}$ of the excited $3d^5(^6S)7p(^5P^o)$ state of Fe~III 
 exhibiting huge Seaton resonance at 1.73 Ry, the transition energy
 for a PEC from the ground $3d^5(^6S)$ to excited $3d^44p(^6P^o)$ state of 
 the core ion Fe~IV.
 Right: $\sigma_{PI}$ of two series of excited states of Fe~III - a) 
$3p^54p(^5P^o)$, b) $3p^55p(^5P^o)$, and c) $3p^58s(^7S)$, d) $3p^59s(^7S)$, 
 e) $3p^510s(^7S)$, f) $3p^511s(^7S)$ illustrating that regardless of the 
state, the energy position (pointed by arrow) of a Seaton resonance remains 
 the same (Nahar 1996).
 }
 \end{figure}

\subsection{Resonances at and near ionization threshold due to relativistic
channels}

Relativistic effects are important when electrons are moving very fast due 
to high nuclear charge of a heavy or high Z element and highly charged ion,
such as, Fe~XXV. Relativistic effects break the resonances in to fine 
structure components causing narrower and larger number of resonances.
However, we have found that fine structure effect can introduce strong 
resonances at and near ionization threshold region of $\sigma_{PI}$ for 
low to medium high Z elements (e.g. for P~II by Nahar et al 2017) but 
not significant for the rest of the energy region.
Figure 5 presents $\sigma_{PI}$ of $2s2p^3(^5S^o)$ of carbon like Fe~XXI 
demonstrating the relativistic fine structure effect introducing strong 
resonances in the near threshold region not allowed in LS coupling 
approximation (Nahar 2008). In LS coupling the state $2s2p^3(^5S^o)$ 
photoionizes only to $2s2p^2(^4P)\epsilon d(^5P)$ or 
$2s2p^2(^4P)\epsilon ns(^5P)$ and 
leave the core ion at $2s2p^2(^4P)$ state. It means that $\sigma_{PI}$ is zero
below the $^4P$ state (pointed by an arrow in Figure 5). However, in reality
resonances can form below the $^4P$ threshold by the fine structure 
couplings effect. The fine structure channels with the excited core ion,
$2s2p^2(^4P)nd,ns(^5P_{1,2,3})$, are allowed to couple with those of 
the ground levels $2s^22p(^2P^o_{1/2,3/2})\epsilon p(^{1,3}P_{1,2,3})$. 
This leads to radiationless autoionization below the $^4P$ threshold and 
introduction of the narrow and high peak resonances almost zero background 
cross section (Nahar 2008). 
These resonances have found to make significant contribution to low
temperature plasmas.

\subsection{Resonant enhancement due to $\Delta n$=1}

The wavefunction expansions considered by the OP included core ion states 
from the n-comlex of the outer electron, i.e. $\Delta n$ =0. It was assumed 
that the high lying states i) are not common in astrophysical plasmas of 
interest and ii) have weak couplings to lower ones causing weaker resonances.
However, it included the rise in the decaying $\sigma_{PI}$ at an inner shell 
ionization threshold of the core ion. The later studies (e.g. for oxygen ions 
Nahar 1998) found that resonances due to high lying the core ion states 
with $\Delta$n=1 are much stronger with high peaks than those with 
$\Delta$n=0 states. Hence without their consideration, integrated cross 
section can be underestimated.

  \begin{figure}
  \centering
   \includegraphics[width=3.8in,height=2.00in]{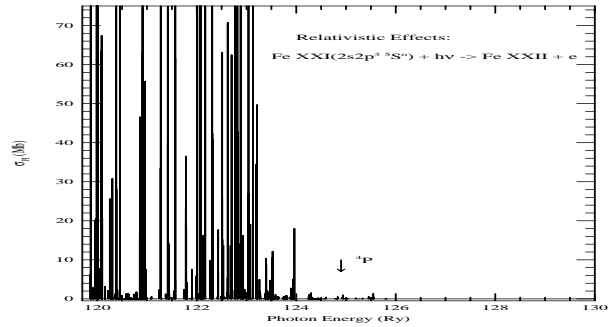}
 \caption{
 $\sigma_{PI}$ of the excited $2s2p^3(^5S^o)$ state of Fe~XXI showing narrow, 
 high peak resonances below the first excited core ion threshold $2s2p^2(^4P)$ 
 (pointed by arrow) formed by fine structure couplings (Nahar 2008).
 }
 \end{figure}

Figure 6 presents $\sigma_{PI}$ of the excited $2s^22p3d(^3D^o)$ state of 
Fe~XXI obtained using (a) a 29-CC wavefunction expansion that includes 
$\Delta n$ = 0,1 states (Nahar 2008a) and (b) a 8-CC expansion which includes 
$\Delta n$ = 0 states only (Lou and Pradhan 1989). 
The first complex of narrow Rydberg resonances (with arrow at n=2) belong 
to excitations of the core ion to $\Delta n$=0 states, that is to states with 
n=2. They are less prominent than those arising from n=3 states in the higher 
energy region where the features are more extensive and stronger spanning over 
a larger energy range. The latter can be divided into two groups: 
the broader Seaton resonances in the photon energy range of 73 to 82
Ry by 8 possible PECs and series of narrow Rydberg resonances belonging to 
various states of n=3 complex in the rest of the high energy region.
The  high energy background has also enhanced considerably. 
Such prominent effects were not unexpected for Fe~XXI because 
of the large energy gap of about 50 Ry between n=2 and 3 states. 
The prominence can be explained by the larger radiative decay rates for n=3 
states.

\subsection{Convergence of enhancement due to Resonances}

Photoionization of Fe~XVII was studied to determine the impact of 
$\Delta n=$2 excitations in the core ion. The $\sigma_{PI}$ of the ion 
contributes significantly to the opacity of the plasma
near the boundary between radiative and convection zones of the sun.
$\sigma_{PI}$ with a wavefunction of 60 fine structure levels (30 LS 
states) with $\Delta n$=0,1 showed considerable amount of photoabsorption 
(Nahar et al 2011) (e.g. $\sigma_{PI}$ in black for the $2s^2p^53d(^1D^o)$
state of Fe~XVII in Figure 7) not seen with those using a 2-CC 
wavefunction (Scott, TOPbase, blue in the figure). The enhancements in 
$\sigma_{PI}$ from 60-CC increased the Rosseland mean opacity for Fe~XVII 
by 20\% from that of 2-CC and explained the possible reason for discrepancy 
with the measured values (Bailey et al 2015). 

The latest study of $\sigma_{PI}$ for Fe~XVII (Nahar and Pradhan 2016) used 
a larger wavefunction expansion of 99 states covering the n=4 states 
($\Delta~n$=0,1,2) (red in Figure 7)).
Arrows point the energy positions for the highest n=2, 3, and 4 states.
We note no significant enhancement from n=4 excitations,  the resonances 
were much weaker and negligible enhancement in the background cross 
section (red) compared to those due to n=3 states (black). 
These demonstrate that a convergence of resonant contributions from core
excitations to higher n states has reached. The narrow resonances close to 
the ionization threshold in the black curve of 30-CC are from fine structure 
couplings of relativistic BPRM method. Such convergence with $\Delta n$ =2
excitations in the core ion was also noted in $\sigma{_PI}$ for Fe~XXV 
(Nahar et al 2001). 
 
 \begin{figure}
 \centering
 \includegraphics[width=4.0in,height=2.00in]{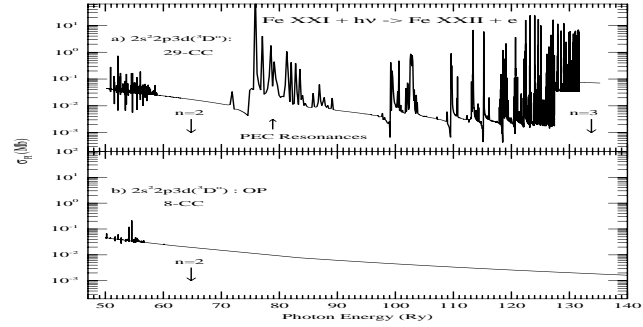}
\caption{
Effect of core ion excitations on $\sigma_{PI}$ of the $2s^22p3d(^3D^o)$
state of Fe~XXI to $\Delta n=0$ states (8-CC, Lou and
Pradhan 1989) and to $\Delta n=1$ states (29-CC, Nahar 2008a)
Comparison shows i) the wide Seaton resonances in the energy range of 73 - 
82.5 Ry, ii) high peak resonances by stronger correlation effect, and iii) 
enhancement in the background at higher energy by the excitations to n=3 states.
}
\end{figure}


\subsection{Resonances below K-Shell ionization}

Resonances below the last ionization threshold, the K-shell threshold, may
show multiple ionization effect. Typically resonances become weaker
and background will drop with higher photon energy. However, resonances
below the K-shell, tend to rise again due to the higher 1s-2p excitation rates.
Their peaks can be orders of magnitude higher than the rise in K-shell 
ionization threshold and the background, such as, over 2 orders of magnitude 
for Fe and over 3.5 orders for AU as presented in Figure 8 (Pradhan et al. 
2009, Lim et al. 2012). Figure 8 shows these resonances added to the background 
cross sections of the atoms available at the website of NIST (NIST2). 
The complex of resonances in Figure 8 correspond to allowed K$_{\alpha}$ 
(1s-2p) transitions of all ionization stages from Ne- to He-like ion following 
K-shell ionization. In the Auger process, where an L-shell electron drops 
down to fill the K-shell vacancy, a K$_{\alpha}$ photon is emitted. This 
photon can knock out an L-shell electron creating an additional hole there. 
Continuation of such process to higher levels can lead to Coster-Kronig 
cascading, that is, multiple ionization stages as photons and electrons are
ejected out. However, such cascade can occur if the photon energy matches to 
those of the resonances presented in the figure (Pradhan et al 2009). The 
enhanced resonant photoionization can result in resonant K$_{\alpha}$ 
fluorescence below the K-shell ionization of each ionization stages of 
Ne-like to He-like ions and was observed by Vinko et al. for Al (2012) and 
interpreted by Nahar and Pradhan (2015).

\section{Accuracy and Completeness}

 \begin{figure}
 \centering
  \includegraphics[width=4.2in,height=2.00in]{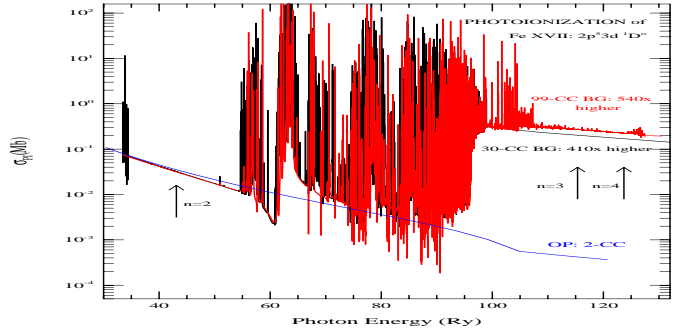}
\caption{
Comparison of $\sigma_{PI}$ of the excited $2s^22p^53d\ ^1D^o$ state of 
Fe~XVII from the three calculations, i) 2-CC in blue (scott), ii) 30-CC 
(60 fine structure levels) in
black (Nahar et al. 2011) and iii) 99--CC in red (Nahar and Pradhan 2016).
The high peak strong resonances and huge background enhancement by orders of
magnitude are due to $\Delta n$=1 states in the wavefunction.
The weaker resonances of n=4 excitations indicate the convergence of resonant
contributions with higher n. The arrows indicate the energy limits for the
n=2,3,4 states of the core ion.
}
\end{figure}

Both accuracy and completeness are needed for calculations of photoionization
cross sections for precise prediction in features, benchmarking 
with high precision experiment, and applications for solving problems. 
Accuracy depends on the number of states included in the 
wavefunction expansion, configurations included for accurate 
representation of correlation effect and number of points for delineation
f resolution 
of the resonances. Completeness means obtaining cross sections for large 
number of bound states. These have been objectives of both the OP and the IP. 
An example of accuracy is given in Figure 9 presenting $\sigma_{PI}$ of 
the complex ion P II where the combined resonant features observed in the 
sophisticated set-up of Advanced Light Source in Berkeley National Lab 
(panel a) are being identified with theoretical calculations using close 
coupling approximation and R-matrix method in panels b-f (Nahar et al 2017). 
Comparison shows that the observed features are reproduced in theory and
that belong to those of the ground and low lying metastable levels.

\section{Conclusion}

The extensive study of atomic photoionization carried out under the Iron
Project and the Opacity Project has established number of characteristic 
features of the process. They have critical impact on the astrophysical 
applications, such as, determination of the plasma opacity, elemental 
abundances, ionization fractions. Detailed study of the process with high 
accuracy and compute large scale atomic data for completeness are necessity
for precise analysis of astrophysical spectra and in other applications. 


 \begin{figure}
 \centering
  \includegraphics[width=4.0in,height=2.00in]{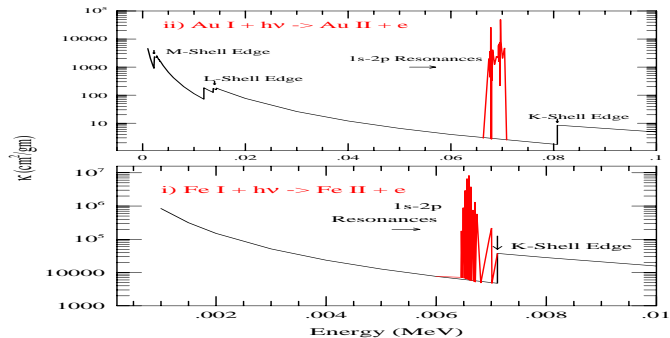}
\caption{High peak resonances below the K-shell ionization threshold
 of i) iron and ii) gold arising from K$_{\alpha}$ (1s-2p) transitions.
 The resonances correspond to all allowed K$_{\alpha}$ transitions of the
 Ne- to He-like ions as the element go through multiple ionization stages
 at a resonant energy (Pradhan et al 2009).
 }
\end{figure}

\acknowledgements Partial support by DOE grant DE-FG52-09NA29580 and NSF grant
AST-1312441. Computations were carried out at
the Ohio Supercomputer Center

\begin{figure}
 \centering
  \includegraphics[width=4.5in,height=2.40in]{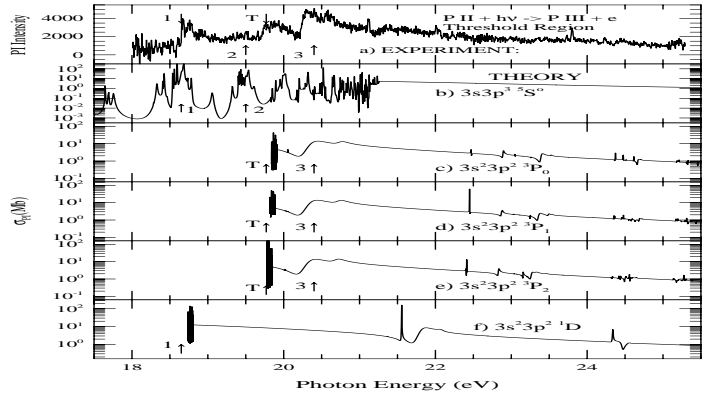}
\caption{
Benchmarking of $\sigma_{PI}$ of P~II for accuracy: (a) the combined features
observed in experiment, (b-f) theoretical
identification of observed features (1,2,3,T) generated by the (b) excited
level $3s3p^3(^5S_2^o)$, (c) ground level $3s^2 \ 3p^2 (^3\mbox{P}_{0})$ and
(d,e,f) the three excited levels $3s^23p^2 (^3P_{1,2})$, $3s^23p^2 (^1D_{2})$.
Agreement is seen in reproducing the observed features (Nahar et al 2017)}
\end{figure}


\begin{thebibliography}{}
\bibitem[{{A}(4)}]{ags1} Asplund M, Grevesse N, Sauval J, arXiv 2004;astro-ph/0410214v2
\bibitem[{{A}(4)}]{ags2} Asplund M, Grevesse N, Sauval J, Scott P, Annu. Rev.
Astron. Astrophys. 47, 481¿522 (2009)
\bibitem[{{A}(4)}]{baileyetal} Bailey et al. (22 authors) Letter, Nature 517, 
56 (2015)
\bibitem[{{A}(4)}]{betal87}  Berrington KA, Burke PG, Butler K, et al, J. Phys. 
B 20, 6379 (1987)
\bibitem[{{A}(4)}]{betal95} Berrington KA, Eissner W, Norrington PH, Comput. Phys. 
Commun. 92, 290 (1995)
\bibitem[{{A}(4)}]{br75}  Burke PG, Robb WD, Adv. At. Mol. Phys. 11, 143-214 (1975)
\bibitem[{{A}(4)}]{ss} Eissner W, Jones M, \& Nussbaumer H, Comput. Phys. Commun. 8,
270-306 (1974)
\bibitem[{{A}(4)}]{gu08} Gu MF, Can. J. Phys. 86, 675¿689 (2008)
\bibitem[{{A}(4)}]{ip} Hummer, D.G., Berrington, K.A., Eissner, W., et al, Astron. 
Astrophys.279, 298-309 (1993)
\bibitem[{{A}(4)}]{letal12} Lim S, Montenegro M, Pradhan AK, Nahar SN, Chowdhury E,
Yu Y, World Congress on Med. Phys. Biomed. Eng.,
IFMBE Proceedings 39, p. 2248 (Ed. M. Long, Springer, 2012)
\bibitem[{{A}(4)}]{lp89}  Luo D, Pradhan AK,  J Phys B 22, 3377¿95 (1989)
\bibitem[{{A}(4)}]{opserver} OPSERVER: Mendoza C, Seaton MJ, Buerger P, et al. 
MNRAS 378, 1031 (2007). Data at http://opacities.osc.edu.
\bibitem[{{A}(4)}]{snn96} Nahar SN, Phys. Rev. A 53, 1545-1552 (1996)
\bibitem[{{A}(4)}]{snn98} Nahar SN, Phys. Rev. A 58, 3766 (1998)
\bibitem[{{A}(4)}]{snn08} Nahar SN, J. Quant. Spec. Rad. Transfer 109, 2731-2742 (2008)
\bibitem[{{A}(4)}]{snn08a} Nahar SN, J.  Quant. Spec. Rad. Transfer 109, 2417-2426 (2008)
\bibitem[{{A}(4)}]{snn1} Nahar SN, unpublished, Resulrs obtained using the wavefunction 
of Nahar SN, Pradhan AK, 437, 345 (2005) 
\bibitem[{{A}(4)}]{netal03} Nahar SN, Eissner W, Chen GX, Pradhan AK. 
A\&A 408, 789-801 (2003)
\bibitem[{{A}(4)}]{netal17} Nahar SN, Hern\'andez EM, Hern\'andez L, et al. 
JQSRT 187, 215-223 (2017)
\bibitem[{{A}(4)}]{netal10} Nahar SN, Montenegro M, Eissner W, Pradhan AK, Phys.Rev. 
A 82, Brief Rep 065401 (2010)
\bibitem[{{A}(4)}]{npm11} Nahar SN, Pradhan AK, Montenegro K, in {\it Simulations 
in Nanobiotechnology}, CRC Press - Taylor \& Francis group, Chap 9, 
p.305-330 (2011)
\bibitem[{{A}(4)}]{np94} Nahar SN, Pradhan AK, J. Phys. B 27, 429 (1994)
\bibitem[{{A}(4)}]{np05} Nahar SN, Pradhan AK, Astron. Astrophys 437. 345 (2005)
\bibitem[{{A}(4)}]{netal11} Nahar SN, Pradhan AK, Chen GX, Eissner W, Phys. Rev. A 83, 
053417 (2011)
\bibitem[{{A}(4)}]{npz01} Nahar SN, Pradhan AK, Zhang HL, Astrophys. J. Suppl. 133, 
255 (2001)
\bibitem[{{A}(4)}]{np15} Nahar SN, Pradhan AK, JQSRT 155, 32-48 (2015)
\bibitem[{{A}(4)}]{np16}  Nahar SN, Pradhan AK, Phys. Rev. Lett. 116, 235003 (2016)
\bibitem[{{A}(4)}]{nist2} NIST2: http://physics.nist.gov/PhysRefData/Xcom/html/xcom1.html
\bibitem[{{A}(4)}]{norad} NORAD atomic data website: http://norad.astronomy.ohio-state.edu
\bibitem[{{A}(4)}]{aas} Pradhan, A.K. \& Nahar S.N. in {\it Atomic Astrophysics and
Spectroscopy} (AAS) (Cambridge University press, 2011)
\bibitem[{{A}(4)}]{petal09} Pradhan AK. Nahar SN, Montenegro N, et al J. Phys. Chem. 
A 113, 12356 (2009)
\bibitem[{{A}(4)}]{pz97} Pradhan AK \& Zhang HL, J. Phys. B 30, L571-L579 (1997)
\bibitem[{{A}(4)}]{rm79} R.F. Reilman and S.T. Manson, Astrophys. J. Suppl. {\bf 40},
815 (1979)
\bibitem[{{A}(4)}]{stb90} Sakimoto K, Terao M, \& Berrington K.A, Phys. Rev. A 42,
291-295 (1990)
\bibitem[{{A}(4)}]{sb92} Sawey PMJ, Berrington KA, J. Phys 8: Al. Mol. Opt. Phys
25 (1992) 1451-1466 
\bibitem[{{A}(4)}]{scott} Scott MP; unpublished, data in TOPbase
\bibitem[{{A}(4)}]{mjs} Seaton, M.J., J. Phys. B, 6363-6378 (1987)
\bibitem[{{A}(4)}]{op} The Opacity Project Team. {\it The Opacity Project}, Vol 1,
1995, Vol. 2, 1996, Institute of Physics
\bibitem[{{A}(4)}]{mjs1} Seaton MJ, unpublished, data available at TOPbase
\bibitem[{{A}(4)}]{setal10} Simon MC, Schwarz M, Epp SW, et al, J. Phys. B: At. Mol. 
Opt. Phys. 43 065003 (2010)
\bibitem[{{A}(4)}]{tipbase} TIPbase: $http://cdsweb.u-strasbg.fr/tipbase/topbase.html$
\bibitem[{{A}(4)}]{topbase} TOPbase: $http://cdsweb.u-strasbg.fr/topbase/topbase.html$
\bibitem[{{A}(4)}]{vetal12} Vinko SM, Ciricosta O, Cho BI, et al., Nature 482, 59 (2012)
\bibitem[{{A}(4)}]{ys87} Yu Y. \& Seaton, M.J. J. Phys. B 20, 6409-6429 (1987)
\bibitem[{{A}(4)}]{znp99} Zhang HL, Nahar SN, Pradhan AK. J. Phys. B 32, 1459-1479 (1999)
%
\end{thebibliography}


\end{document}